\documentstyle[prl,aps,amsmath,graphicx,epsfig,psfig,
               pstricks]{revtex}

\newpsobject{showgrid}{psgrid}{subgriddiv=1,griddots=10,gridlabels=7pt}

\DeclareMathOperator{\I}   {i}
\newcommand{\diff}[2]{\partial_#2 #1}
\newcommand{\abs}  [1]{\lvert#1\rvert}
\newcommand{\nabn}{\nabla^2}
\newcommand{\eno}  [1]{\mbox{\!(\ref{#1})}}

\begin{document}
\draft
\twocolumn
\title{Self-organized stable pacemakers near the onset of birhythmicity}
\author{Michael Stich, Mads Ipsen, and Alexander S. Mikhailov}
\address{Fritz-Haber-Institut der Max-Planck-Gesellschaft, Faradayweg
4-6, D-14195 Berlin, Germany}
\date{\today}
\maketitle

\begin{abstract}
General amplitude equations for reaction-diffusion systems
near to the soft onset of birhythmicity described by a
supercritical pitchfork-Hopf bifurcation are derived. Using
these equations and applying singular perturbation theory, we
show that stable autonomous pacemakers represent a generic
kind of spatio\-temporal patterns in such systems. This is
verified by numerical simulations, which also show the
existence of breathing and swinging pacemaker solutions. The
drift of self-organized pacemakers in media with spatial
parameter gradients is analytically and numerically
investigated.
\end{abstract}




Oscillatory reaction-diffusion systems, such as the
Belousov-Zhabotinsky (BZ) chemical reaction, exhibit a rich
variety of nonlinear wave patterns.  The first complex pattern
discovered in this reaction was the target pattern, where
concentric waves were emitted by a pacemaker representing a
periodic wave source~\cite{Zaikin70}.  Subsequently, similar
target structures have been observed in many other chemical,
physical, and biological
systems~\cite{JakRo90,AssStein93,NaSaSa89,Lee97}. A simple
theoretical explanation of target patterns in oscillatory
chemical systems is that their pacemakers are created by
impurities which increase the local oscillation frequency in
the medium~\cite{TyFi80}. Most of the target patterns seen in
the BZ~reaction are indeed caused by small local
inhomogeneities, such as dust particles. A general question is
whether self-organized target patterns, representing an
intrinsic dynamical property, are also possible in
reaction-diffusion systems.  Examples of stable autonomous
pacemakers with localized or extended wave patterns are indeed
known for several reaction-diffusion
models~\cite{Vasiliev86,Mik92,KawCom92,Saka92,Kokubo94,DeiBra94,ZhDoEp95,OhHaKo96,RoZhEp97,NekShe98}.

The aim of the present Letter is to show that stable autonomous
pacemakers with extended wave patterns represent a {\em generic\/}
pattern-forming object in oscillatory reaction-diffusion systems near
the onset of {\em bi\-rhyth\-micity}.  Birhythmicity, the coexistence
of two stable limit cycles corresponding to uniform oscillations with
different frequencies, is possible in various
systems~\cite{DeGold82,AlEp83,LaHu85,PerezI98}, including glycolytic
oscillations~\cite{Gold96} and the photosensitive BZ
reaction~\cite{KrKu90}. Here, we derive two coupled amplitude
equations yielding the normal form of such a dynamical system near a
supercritical pitchfork-Hopf bifurcation which leads to birhythmicity.
Using singular perturbation theory, an analytical solution for
autonomous pacemakers is then constructed and its stability is
numerically confirmed. In addition, target patterns with breathing or
swinging pacemakers are observed.  Finally, we show that autonomous
pacemakers can drift under the influence of a parameter gradient and
determine the drift velocity.

The derivation of normal forms for various kinds of reaction-diffusion
systems has recently been discussed by one of the
authors~\cite{Trans99,IpsKraSo00}. Here, we focus our attention on
systems near to a supercritical pitchfork-Hopf bifurcation, where a
real uniform eigenmode and a pair of complex conjugate uniform
eigenmodes start to grow simultaneously. This implies that either a
stable small-amplitude limit cycle becomes unstable and gives rise to
two stable limit cycles or a pair of stable and unstable limit cycles
of small amplitude emerges near to a stable limit cycle. At least
three species are needed to realize this bifurcation.

Using the approach described in~\cite{Trans99}, we have derived the
normal form of this bifurcation for a general reaction-diffusion
system~\cite{IpsStiMik01}. The normal form is given by
\begin{subequations}
  \begin{align}
    \label{eq:ScaledEqn-A}
    \diff{A}{t} &=  A - (1 + \I\alpha )\abs{A}^{2} A +
    (1+\I\beta )\nabn A + (1 - \I \epsilon)A z, \\
    \label{eq:ScaledEqn-B}
    \tau \diff{z}{t} &= \sigma - \gamma \abs{A}^2 + z -\nu z^3
    + \lambda^{2}\nabn z,
  \end{align}
  \label{eq:ScaledEqn}
\end{subequations}
The system~\eno{eq:ScaledEqn} represents a complex
Ginzburg-Landau equation (CGLE) for a supercritical Hopf
bifurcation~\eno{eq:ScaledEqn-A} which is coupled to an
equation describing an imperfect pitchfork
bifurcation~\eno{eq:ScaledEqn-B}. Here $A$ is the complex
oscillation amplitude and $z$ is the amplitude of the slow
real mode. The coefficients $\tau$ and $\lambda$,
respectively, are the ratios of the characteristic time and
length scales of the real and the oscillatory mode. The
parameter $\epsilon$ specifies the frequency shift of the
oscillatory mode due to coupling to the real mode, $\gamma $
characterizes the strength of the feedback from the
oscillatory to the real mode, and $\nu$ determines the
nonlinear saturation of the real mode. When $\sigma =0$,
Eq.~\eno{eq:ScaledEqn-B} describes a supercritical pitchfork
bifurcation, whereas $\sigma \neq 0$ corresponds to an
imperfect pitchfork (or ``cusp''~\cite{Kutz95})
bifurcation. Only positive parameters $\gamma ,\nu$, and
$\sigma $ will be considered in this Letter. We assume that uniform
oscillations are modulationally stable in this system, i.e.\
the Benjamin-Feir-Newell condition $1+\alpha \beta >0$ is
satisfied, and that the waves have positive dispersion, i.e.\
$\beta -\alpha >0$.

It can be seen that Eqs.~\eno{eq:ScaledEqn} have solutions
corresponding to two uniform stable limit-cycle oscillations
with frequencies $\Omega _{1}$ and $\Omega _{3}$ and to an
unstable limit cycle with frequency $\Omega _{2}$.  The
frequencies are given by $\Omega _{1,2,3}=\alpha +\left(
\alpha +\epsilon \right) z_{1,2,3}$ where $z_{1,2,3}$ are the
real roots of the equation $\nu z^{3}-(1-\gamma )z+\gamma
=\sigma $. When $\alpha +\epsilon <0$ (which is the case that
we chose in the simulations), the smallest root $z_{1}$
corresponds to the most rapid oscillations, i.e.\ $\Omega
_{3}<\Omega _{2}<\Omega _{1}$.

In addition to these uniform oscillations, the system
described by Eqs.~\eno{eq:ScaledEqn} may have stable
nonuniform solutions representing self-organized
pacemakers. To create a pacemaker, a sufficiently strong local
perturbation should be applied to the state comprised of
uniform oscillations with the lower frequency $\Omega_{3}$, so
that a small core region is formed where oscillations have a
higher frequency. Inside this region, the variable $z$ is
close to $z_{1}$, whereas outside, it is near to $z_{3}$. The
core starts to send out waves and hence a pacemaker is
created. The core expands ($\gamma >\sigma$ is a necessary
condition for this, otherwise it will contract) and the
frequency and the wavenumber of the emitted waves slowly
increase at the same time.  In turn, this leads to a decrease
of the oscillation amplitude of emitted waves. This amplitude
controls the propagation velocity of the front, representing
the boundary of the expanding core. When a critical wavenumber
is reached, the front velocity becomes zero and a stationary
pacemaker is formed.

We have constructed an analytical solution for stationary
pacemakers in the one-dimensional
system~(\ref{eq:ScaledEqn}). The phase $\phi$ and amplitude
$\rho $ are introduced by \mbox{$A=\rho\exp[-\mbox{i}(\Omega_3
t+\phi)]$}. Then $\rho$ is adiabatically eliminated using
\mbox{$\rho ^{2}\approx 1+z-(\nabla \phi )^{2}+\beta \nabla
^{2}\phi $}. The phase equation approximation~\cite{Kuramoto}
is valid for smooth phase perturbations and $\nu \gg
1-\gamma$.  Assuming that the characteristic length scale of
the real mode, determining the front width, is much shorter
than the characteristic length scale of the oscillatory
subsystem (i.e.\ that $\lambda \ll 1$), we apply singular
perturbation theory to this problem. The derivation will be
published separately~\cite{StIpMi2} and only selected results
are reported in this Letter.

\begin{figure}[htbp]
  \noindent
  \begin{pspicture}(0,0)(8.5,3.3)
    \rput[bl](0.0,0.0){%
      \includegraphics[width=8.65cm]{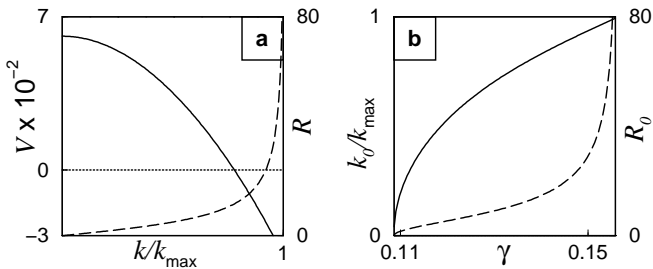}
      }
  \end{pspicture}
  \caption{%
    (a) Dependence of the front velocity $V$ on the wavenumber
    $k$ of emitted waves (solid line) and dependence of $k$ on
    the core radius $R$ (dashed line, plotted as $R$ {\em vs.}
    $k$). (b) The wavenumber $k_{0}$ of the waves emitted by a
    stationary pacemaker (solid line) and the corresponding
    radius $R_{0}$ (dashed line) as functions of the coupling
    coefficient $\gamma$. The parameters are $\alpha =1.4$,
    $\beta=2.3$, $\epsilon =-2.1$, $\lambda=1$, $\tau =5$,
    $\nu =20$, $\gamma=0.13$, $\sigma=0.1$.}
  \label{fig1}
\end{figure}

The velocity of the front $V$ (of the expanding core) depends
on the wavenumber $k$ and is given by
\begin{equation}
V(k)=3\frac{\lambda }{\tau }\sqrt{\frac{\nu }{2}}\widetilde{z}_{2}(k),
\label{V}
\end{equation}
where $\widetilde{z}_{2}(k)$ is the middle root of the cubic
equation \mbox{$\nu z^{3}-\left( 1-\gamma +\gamma a\right)
z+\gamma (1+az_{3}-k^{2})=\sigma$}, with \mbox{$a=\beta
(\alpha +\epsilon )/(1+\alpha \beta )$.} On the other hand, if
the core radius $R$ is known, the wavenumber $k$ of emitted
waves can be analytically found if the condition $\lambda
k\ll1$ is satisfied (cf.~\cite{Mik92}). The inverted
dependence $R(k)$ and the velocity $V(k)$ are displayed in
Fig.~\ref{fig1}(a).

For stationary pacemakers, the front velocity $V$
vanishes. This determines the wavenumber $k_{0}$ of a
stationary pacemaker and thus allows us to find its core
radius $R_{0}$. Analytical solutions for these key properties
have been constructed. We find that
\begin{eqnarray}
k_{0}&=&\sqrt{1-\sigma /\gamma +az_{3}}\,,  \label{k0} \\
R_{0}&=&\frac{1+\alpha \beta }{(\beta -\alpha )\sqrt{k_{{\rm {max}}
}^{2}-k_{0}^{2}}}\,{\rm tan}^{-1}\!\left( \frac{k_{0}}{\sqrt{k_{{\rm {max}}
}^{2}-k_{0}^{2}}}\right),
\end{eqnarray}
where $k^2_{{\rm {max}}} = (\alpha+\epsilon)
(z_{1}-z_{3})/(\beta-\alpha)$.  The frequency $\Omega _{0}$ of
a stationary pacemaker is $\Omega _{0}=\Omega _{3}+(\beta
-\alpha )k_{0}^{2}$. The wavenumber $k_{0}$ and the radius
$R_{0}$ of a stationary pacemaker are shown as functions of
the coupling coefficient $\gamma$ in Fig.~\ref{fig1}(b).

Examining the constructed solutions, we note that generally
$\Omega_{3}<\Omega _{0}<\Omega _{1}$. The frequency $\Omega
_{0}$ of a stationary pacemaker approaches the frequency
$\Omega _{1}$ of rapid uniform oscillations, when the core
radius $R_{0}\rightarrow \infty$ (and $k_{0}\rightarrow
k_{\max }$). On the other hand, when the core is small,
$k_{0}$ is small and the frequency $\Omega _{0}$ is close to
$\Omega _{3}$.  Stationary pacemakers exist inside an interval
of the coupling intensity $\gamma$ [see
Fig.~\ref{fig1}(b)]. Our approximate analysis based on
singular perturbation theory is only valid when the core is
not too small, i.e.\ $R_{0}\gg \lambda$.

Some conclusions about the stability of stationary pacemakers
can already be drawn from Fig.~\ref{fig1}(a). Suppose the
radius $R$ has increased above the stationary radius
$R_{0}$. This leads to an increase of the wavenumber $k$ of
emitted waves which, in turn, will make the front velocity $V$
negative. Therefore the front will retreat, decreasing the
radius $R$ back to its stationary value. This argument is,
however, only applicable when the characteristic time scale of
the core evolution is much longer than the time needed for the
wave pattern to adjust to its changes, i.e.\ when $\tau \gg
1$.  Generally, the stability of stationary pacemakers should
be numerically investigated.

\begin{figure}[htbp]
  \noindent
  \begin{pspicture}(0.0,0)(8.5,5.4)
    \rput[bl](0.05,3.05){%
      \includegraphics[width=8.5cm]{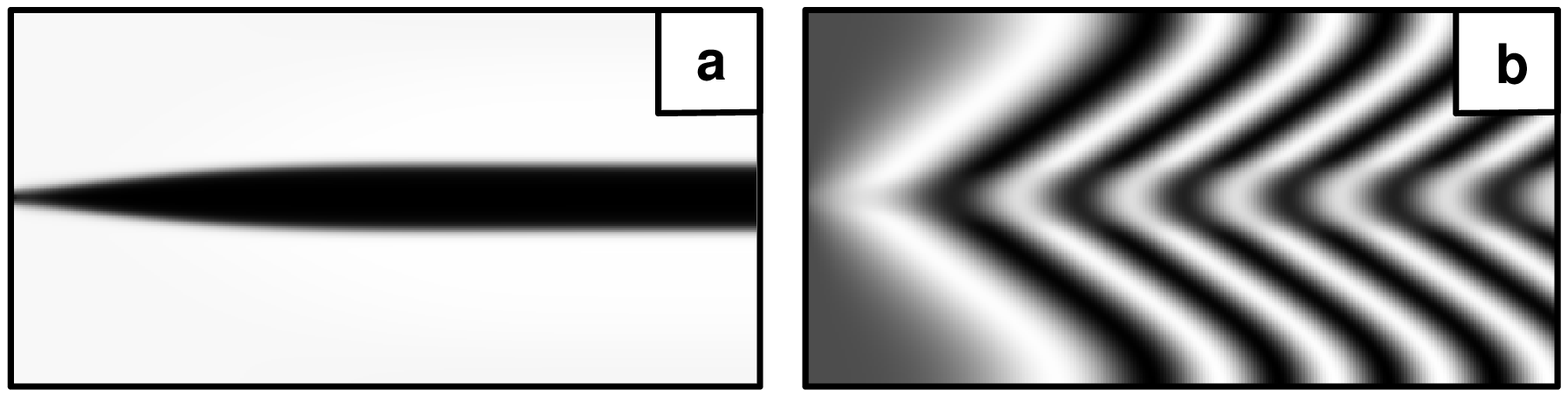}%
      }
    \rput[bl](0.05,0.25){%
      \includegraphics[width=9.1cm]{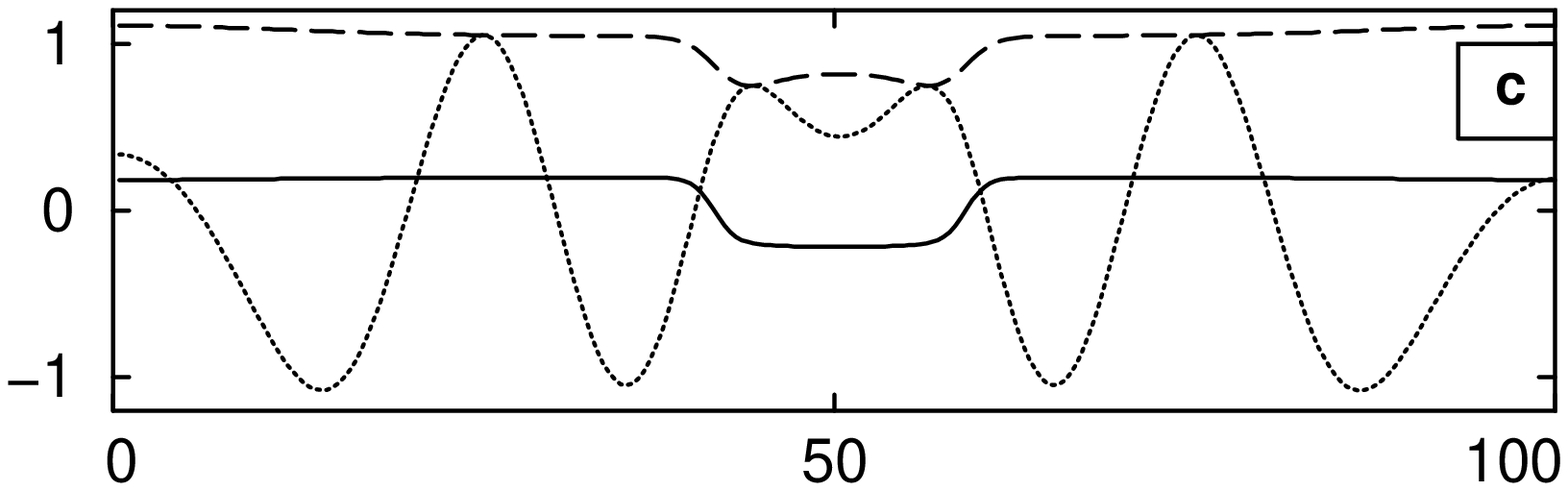}
      }
  \end{pspicture}
  \caption{%
    Development of a stable stationary pacemaker (a,b) and its
    asymptotic profile (c). Frame (a) shows the evolution of
    the real mode amplitude $z$ after an initial
    perturbation. Frame (b) displays the evolution of
    Re$A$. In frame (c) the spatial distribution of the
    variables $z$ (solid line), Re$A$ (dotted line) and $|A|$
    (dashed line) are presented. The system size is $L=100$
    and the time interval is $0<t<500$; the same parameters as
    in Fig.~\protect\ref{fig1}. In our gray-scale plots the
    black and white levels always correspond to the minimum
    and the maximum values of the plotted variable,
    respectively. In the space-time plots, time runs along the
    horizontal axis.}
\label{stable1d1}
\end{figure}

The system described by Eqs.~\eno{eq:ScaledEqn} was integrated
with an explicit Euler scheme where the Laplacian operator was
discretized with a nearest-neighbor approximation. No-flux
boundary conditions were used. Figure~\ref{stable1d1}(a)
displays the evolution of a stationary pacemaker from a small
initial perturbation of the real mode $z$. In the first stage,
the core grows with approximately constant speed. Later, the
growth is terminated and a stationary object is
formed. Figure~\ref{stable1d1}(b) displays the creation and
emission of waves in the oscillatory subsystem. The profile of
the asymptotic stable stationary pacemaker is shown in
Fig.~\ref{stable1d1}(c).

\begin{figure}[htbp]
  \noindent
  \begin{pspicture}(0.0,0)(8.5,4)
    \rput[bl](0.0,0.1){
      \includegraphics[width=8.5cm]{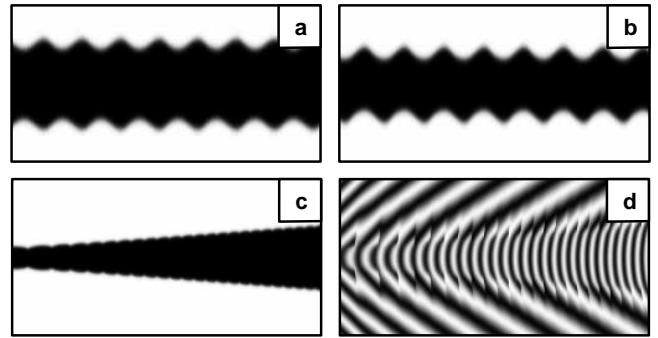}
      }
  \end{pspicture}
  \caption{%
    Breathing (a), swinging (b), and Eckhaus-unstable (c-d)
    pacemakers.  The displayed coordinate and time ranges are
    $\Delta L=50$, $\Delta T=125$ (a-b) and $\Delta L=100$,
    $\Delta T=625$ (c-d). The parameters are $L=100$,
    $\alpha=1.38$, $\epsilon =-3.18$, $\lambda=0.8$, $\nu=83$,
    $\gamma=5.59\cdot 10^{-4}$, $\sigma=3.4\cdot 10^{-4}$, and
    for (a): $\beta=3.0, \tau=0.001$, (b):
    $\beta=2.65,\tau=0.001$, (c-d): $\beta=2.1, \tau=0.025$.}
\label{comb1d1}
\end{figure}

Pacemakers are stable for sufficiently large $\tau$. When
$\tau$ is decreased, numerical integrations show that
stationary pacemakers become unstable. Close to the
instability boundary, stable breathing and swinging pacemakers
were found [Figs.~\ref{comb1d1}(a)--\ref{comb1d1}(b)]. For a
breathing pacemaker, the center remains stationary whereas the
radius oscillates. For swinging pacemakers, the radius stays
approximately constant while the position of the pacemaker
oscillates. Further lowering of $\tau$ leads to the
disappearance of any stable pacemaker solutions.

Similar to plane waves in the CGLE~\cite{JanPum92}, the waves
emitted by a pacemaker may become unstable when their
wavenumber $k_0$ exceeds the Eckhaus wavenumber given by
$k_{{\rm {EH}}}\approx \sqrt{(1+z_{3})(1+\alpha \beta
)(3+\alpha \beta +2\alpha ^{2})^{-1}}$.  Among other effects,
this may lead to the destabilization of a stationary
pacemaker, as illustrated by
Figs.~\ref{comb1d1}(c)--\ref{comb1d1}(d).  Phase singularities
and thus oscillation amplitude defects are periodically
generated at the core boundary, giving rise to a
short-wavelength regime in the core and to a long-wavelength
regime in the periphery. The core gradually grows and
eventually the whole medium is occupied by rapid uniform
oscillations.

In contrast to pacemakers which are created by local
heterogeneities, autonomous pacemakers are not pinned and
their location is determined only by the initial
conditions. Moreover, such self-organized structures are able
to move through the medium when spatial parameter gradients
are present.  Suppose, for example, that the parameter $\gamma
$ varies with a constant gradient $\kappa$, i.e.\
$\gamma(x)=\gamma _{0}+\kappa (x-x_{0})$, where $\gamma _{0}$
is the value of $\gamma$ in the center of the pacemaker. For
sufficiently small gradients ($\kappa R_{0}\ll \gamma _{0}$),
linear perturbation theory can be used. Its application
(see~\cite{StIpMi2}) allows us to determine analytically the
drift velocity $V_{{\rm {D}}}$ as \mbox{$V_{{\rm {D}}}=\kappa
R_{0}\, \partial_{\gamma} V(\gamma_0,k_0)$} where $V(\gamma
,k)$ is the front velocity given by Eq.~\eno{V}.

\begin{figure}[htbp]
  \noindent
  \begin{pspicture}(0.0,0)(8.5,2)
    \rput[bl](0.0,0.1){
      \includegraphics[width=8.5cm]{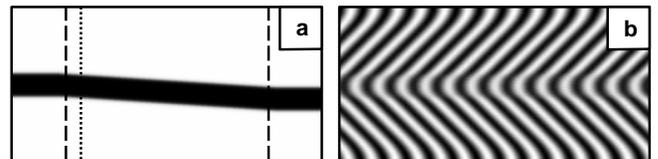}
      }
  \end{pspicture}
  \caption{%
    Drift of a pacemaker. The spatial gradient of the
    parameter $\gamma $ with $\kappa /\gamma _{0}=0.003$ is
    applied inside the time interval indicated by vertical
    dashed lines in frame (a), which shows the evolution of
    $z$ in the time interval $0<t<2\cdot 10^{5}$. The
    pacemaker is drifting in the direction of increased
    $\gamma$. The parameters are $\gamma_0=5.59\cdot 10^{-4}$,
    $\beta=2.3$, $\tau=2$.  The rest are the same as in
    Fig.~\protect\ref{comb1d1}. Frame (b) displays the
    drifting wave pattern within a narrow time interval
    $\Delta T=500$ during the drift, marked by the dotted
    vertical line in frame (a).}
  \label{drift1d1}
\end{figure}

The simulation displayed in Fig.~\ref{drift1d1}(a) was
initiated with a stable stationary pacemaker. After a constant
gradient in the parameter $\gamma$ was introduced, the
pacemaker drifted through the medium in the direction of
increasing $\gamma$. When the gradient was removed, the drift
of the pacemaker terminated and a spatially shifted stationary
pacemaker was recovered. The emission and propagation of waves
persisted during the drift [Fig.~\ref{drift1d1}(b)]. In
addition, the Doppler effect led to a small increase of $k$ in
the direction of motion.

In this Letter, we have analytically constructed
self-organized pacemaker solutions in the vicinity of a
pitch\-fork-Hopf bifurcation.  Our numerical investigations
have shown that such self-organized patterns are stable for a
wide range of parameters. To create autonomous pacemakers, a
sufficiently strong local perturbation should be applied to
the state corresponding to stable uniform oscillations.  This
is in contrast to~\cite{RoZhEp97}, where autonomous target
patterns were found near a Hopf bifurcation with a finite
wavenumber and thus uniform oscillations of the medium were
absolutely unstable.  Our approach is also different from the
model~\cite{Saka92,Kokubo94} that was constructed to explain
target pattern formation in electrohydrodynamic convection and
which is based on a Hopf bifurcation of a cellular spatial
structure. On the other hand, Ohta {\em et
al.}~\cite{OhHaKo96} have investigated a two-component
activator-inhibitor model with coexistence of excitable
kinetics and stable uniform oscillations, and reported several
different kinds of autonomous wave sources. The subsequent
numerical studies~\cite{OhtaPC} have, however, shown that
while localized target patterns are stable, target patterns
which extend over the whole medium are unstable in the model
and slowly evolve into uniform oscillations. Stable localized
target patterns are also found in the quintic
CGLE~\cite{DeiBra94}.

Since our analysis is based on general amplitude equations,
the results presented here are valid for any
reaction-diffusion system near a soft onset of birhythmicity
with small-amplitude limit cycles.  In a separate
publication~\cite{IpsStiMik01}, this analysis will be applied
to a particular chemical model system. As in the case of a
Turing-Hopf bifurcation, the results of our analysis based on
the amplitude equations may remain (qualitatively) applicable
even at significant separation from the bifurcation point.
Finally, we note that the physical mechanism responsible for
the stabilization of pacemakers in the considered system
involves a long-range negative feedback, similar to the one
necessary for the formation of stable localized spots in
reaction-diffusion models with fast inhibitor diffusion. Here,
however, an infinite-range inhibition is caused not by
diffusion, but by non-damped propagation of waves emitted from
the core region. The effect of pacemaker drift in systems with
spatial parameter gradients provides a convenient experimental
method to identify self-organized pacemakers and distinguish
them from other target patterns caused by local
heterogeneities in the medium.

The authors thank T. Ohta for an interesting discussion and
H. Engel for bringing Ref.~\cite{KrKu90} to our
attention. Financial support of the Humboldt Foundation
(Germany) is gratefully acknowledged.

\bibliographystyle{prsty}
\bibliography{main}

\end{document}